# Non-local generation of entanglement of photons which do not meet each other

Jürgen Rösch [1], Xian-Min Jing [2], Juan Yin [2], Tao Yang [2], Jian-Wei Pan [1,2]

[1] Physikalisches Institut, Universität Heidelberg, Philosophenweg 12, D-69120 Heidelberg, Germany

[2] Hefei National Laboratory for Physical Sciences at Microscale and Department of Modern Physics, University of Science and Technology of China, Hefei, Anhui 230026, China

We report for the first time in an ancilla-free process a non-local entanglement between two single photons which do not meet. For our experiment we derive a simple and efficient method to entangle two single photons using post-selection technology. The photons are guided into an interferometer setup without the need for ancilla photons for projection into the Bell-states. After passing the output ports, the photons are analyzed using a bell state analyzer on each side. The experimental data clearly shows a non-local interaction between these photons, surpassing the limit set by the CHSH-inequality with an S-value of 2.54 and 24 standard deviations.

In the year 1935 Einstein, Podolsky and Rosen (EPR) published their famous article "Can Quantum- Mechanical Description of Physical Reality Be Considered Complete?" [1, 2] in which they did show the incompatibility between a local realist point of view and the way quantum mechanics looks upon the world. Originally, EPR wanted to show that quantum mechanics in its current form is incomplete - and needs further description which later led to the concept of hidden variables. Since the introduction of Bell's in 1964 many experiments (see e.g. [4–6]) were performed indicating that nature behaves like the mathematical model of quantum mechanics predicts. The first Bell tests have been carried out by Freedman and Clauser [9] who investigated polarization correlations for photons emitted in a cascade process from a calcium source. Experiments which came closer to Bell's' original idea were performed by Aspect and his colleagues [4, 10, 11] where polarization settings where changed within the flight time of the photons and by Weihs et al. [5] who implemented a much stronger condition on locality with truly independent observers where the decision about measurement settings was based on pure statistics.

Creation of entanglement has been demonstrated by various methods which are illustrated in figure 1(a)-1(c). Most of the experiments use a down-conversion source (SPDC) to generate entangled pairs of photons. In this way photons get already entangled at the source where they are created (1a). A technically more challenging method demonstrated by Shih and Alley in [12] is to take orthogonally polarized photons, e.g. one H and one V polarized photon (from preferably independent sources), and overlap them spatially and temporally on a beamsplitter with equal reflectivities (illustration 1b). It is to mention that the processes in figure 1a and 1b require for the two photons (as shown in figure 1a) to interact with each A completely different approach (illustrated in 1c) has been undertaken by Pan et al. in [13] where they use four photons produced by two down-conversion sources prepared in the state

$$|\psi_{1234}\rangle = \frac{1}{2}(|H\rangle_1|V\rangle_2 - |V\rangle_1|H\rangle_2) \otimes (|H\rangle_3|V\rangle_4 - |V\rangle_3|H\rangle_4)$$

Conditional on a projective measurement performed on photon 2 and 3, which are prepared in the antisymmetric $|\Psi^->$-state and which overlap on a beamsplitter, the photons 1 and 4 are projected in that one of the four Bell states that is equal to the state of photons 2 and 3. Using this method to create entanglement of photon 1 and photon 4, two ancilla photons are needed as mediator.

In figure 1(d) a simplified drawing is given which shows the basic principle of our experiment. In contrast to previous setups, our experiment is the first of its kind (ancilla-free process) that creates an entangled Bell State using two individual photons which do not meet each other. The essential experimental setup is illustrated in figure 2. The idea is based on a slightly modified version of the theoretical concept of Pati an Zukowski [21]. Two single photons are guided onto an 8-port interferometer consisting of two beamsplitters with transmission/reflection ratio of 50/50 and a pair of polarizing beamsplitters. For each photon 1 and 2 there is a 50 percent chance to finally reach either one of the detectors D1/D2 on Alice's side and a 50 percent chance to reach either one of the detectors D3/D4 on Bob's side. The electronic circuit connected to the detectors registers the number of coincidences between the detectors D1 and D3, D1 and D4, D2 and D3 and D2 and D4, i.e. coincidences are defined as cases where one photon arrives at Bob's side and one photon arrives at Alice's side. Given these cases, each photon will travel a different path and the individual geometric paths of the photons do not cross each other. When analyzing the photons after the polarizing beamsplitters we can show that an entanglement takes place even when there is no direct "contact/encounter" of one photon with the other. It can be said that there is no possibility of local interaction of the photons similar to the entanglement swapping experiment of Pan et al. [13]. In our experiment we go a step further and present a sinusoidal curve which has visibility high enough to conduct a Bell-type measurement [7] with a convincing violation of the CHSH inequality. For the setup we use a 460 mW single mode Argon-Ion UV Laser beam focused (waist 80 μm) into the center of a 2 mm long beta-barium-borate (BBO) crystal for type-II phase matching. Unwanted laser-fluorescence is minimized with a dispersion prism. Extraordinary and ordinary photons have different velocities and travel along different paths inside the crystal due to the birefringence of the BBO. The resulting walk-off effects are compensated by a combination of a half wave plate and an additional BBO of half the length of the downconversion crystal in each arm. Further polarization optics allow to create a Phi state, namely

$$\Phi^- = \frac{1}{\sqrt{2}}\left(|H\rangle_1|H\rangle_2 + |V\rangle_1|V\rangle_2\right)$$

Since we need in our experiment a high quality single photon source, but not an entangled state of photons, we destroy the entanglement. This can be achieved by filtering out either one, either the HH case or the VV case and can be realized using polarization filters. Hence, the input for the interferometer becomes:

$$\psi_{input} = (|V_1\rangle|V_2\rangle)$$

One key to our experiment is to eliminate any timing information which would make the photons distinguishable. To ensure that both photons enter the interferometer at the same time, we allow path adjustments using an optical trombone prism (prism 2 in figure 2). Also it is important to ensure identical temporal modes inside the interferometer, therefore we introduce two additional optical trombones, prism 1 and prism 3. To check whether the condition of equal path lengths is fulfilled a coincidence measurement at PBS 3 and PBS 4 is performed. As can be seen in figure 3 and 4, we achieve a visibility of 92.3% and 93,4% respectively. Theory would expect a 100% visibility, which is equivalent to a complete breakdown of coincidences for two impinging photons on different sides of the beamsplitter. Practically, this value suffers from small imperfections which are mainly a tribute to slightly imperfect mode-matching. At BS2, Photon 2, which is vertically polarized, has the choice to go to either Alice's side or Bob's side, but its state will in any case be converted to H. Analyzing all pathways which could lead to one of the four possible combinations of coincidences between any of the detectors on Alice's side and one on Bob's side shows:

$$\frac{1}{\sqrt{2}}(a_{1H}^* + ib_{1H}^*) \otimes \frac{1}{\sqrt{2}}(a_{2V}^* + ie^{i\varphi}b_{2V}^*)|0\rangle$$
$$= \frac{1}{2}(a_{1H}^* a_{2V}^* + i^2 e^{i\varphi} b_{2V}^* b_{1H}^* + ie^{i\varphi} a_{1H}^* b_{2V}^* + ia_{2V}^* b_{1H}^*)|0\rangle$$

Therefore photons are projected into a maximally entangled state:

$$|\Psi\rangle = \frac{1}{\sqrt{2}}\left(|H\rangle_1|V\rangle_2 + e^{i\varphi}|V\rangle_1|H\rangle_2\right)$$

where we can arbitrarily change φ with a nanometer stepsize piezo and therewith have full control over the phase. It is important to note, that for all coincidences the two photons do not encounter each other. The argument is independent of the dualism of light and works in the particle framework as well as in the framework of wavepackets. To test for non-local interactions we make use of the inequality first derived by Clauser, Horne, Shimony, and Holt (CHSH) [7, 8] which is more suitable for measurements with polarized photons. This variant of Bell's inequality can be expressed by:

$$S \leq 2$$

where

$$S = E_1(\alpha,\beta) + E_2(\alpha',\beta) + E_3(\alpha,\beta') - E_4(\alpha',\beta')$$

is the so called Bell-parameter with

$$E_x = \frac{N(\alpha,\beta)_{++} + N(\alpha,\beta)_{--} - N(\alpha,\beta)_{+-} - N(\alpha,\beta)_{-+}}{N(\alpha,\beta)_{++} + N(\alpha,\beta)_{--} + N(\alpha,\beta)_{+-} + N(\alpha,\beta)_{-+}}.$$

In each run of the four subexperiments of the CHSH-test the polarizers take one of two possible settings. The measured E-values are coincidence expectation values for different settings of the half wave plates in front of the polarizing beamsplitters PBS 5 and PBS 6. A comparison of classically expected coincidences with quantum mechanically expected coincidences [7, 8] shows that under classical assumptions an absolute value for the Bell-parameter is expected which is not larger then 2 whereas the laws of quantum mechanics allow for S to reach values up to $2\sqrt{2} \approx 2.81$.

We took data for the following settings of the half wave plates: α = 22.5° (⇒ 45°), β = 45° (⇒ 90°), α = 67.5° (⇒ 135°), β' = 0° (⇒ 0°).

Parameters for the angles of the half wave plates are chosen such that the violation is supposed to be at its maximum value. Hence, with the data from table 1 we obtain as the result for the Bell parameter S a value of 2.54 (σ = ±0.023). Therefore the classical value of 2 is violated with 24.08 standard deviations. Applying the CHSH-inequality implicates the 'fair sampling' assumption [17]. For light intensities on the single photon level, highly sensitive photodetectors are needed in the experiment (we used SPCM-AQR13 operated in Geiger-mode). At the current level of technology such detectors have typical quantum efficiencies of 74% at our wavelength (702 nm) and as consequence not all of the created photons can be registered. 'Fair sampling' assumes that the registered and analyzed photons are a fair sample of all emitted photons and do behave the same way as the uncollected photons, i.e. we assume that the photons registered by the detectors are a representative sample of all photons created [10].

Ultra high stability has been a major challenge for our experiment. The interferometer's geometry has carefully been designed in a way which minimizes unwanted changes in path length due to disturbances. Laboratory air-conditioning did show a strong effect on the data. A thick acrylic glass cover was constructed protecting the interferometer from unwanted air-flow and temperature drifts. The cover was designed to allow control for all necessary angular settings from the outside for all four subexperiments. After each subexperiment we performed a measurement in the 45° basis to confirm that we are still in the $\psi^-$-state. Compactness of the setup, highly stable pillar posts and the custom built glass cover enabled us to keep the phase stable for several minutes. Detection events were registered by a self-developed constant fraction discriminator and self-developed counting card in combination with NIM-electronics for the logic.

In summary, our work supports that creation of entanglement in between two photons is possible, even if there is no direct local interaction between the involved photons. The possible choices of the paths the photons can take can be seen as the underlying cause for the observed non-local entanglement-generation.

It can be argued that 50 percent of all produced photons that enter the interferometer meet each other (25 percent at PBS 3 and 25 percent at PBS4 assuming no other losses). In all these cases the photons go to the same side, either both to Alice or both to Bob such that no coincidences occur. However, these events are filtered out (post selection) and are not registered by the electronics. In different words, 50 percent of all incoming photons meet but these are not the cases for which we prove entanglement. As shown by Popescu, Hardy and Zukowski in [20], post selection does not prevent us to perform entanglement verification.

It is to mention that the observed two-photon interference is deeply rooted in the fundamentals of quantum mechanics and the question for hidden variables. The practical applications of the EPR-principle extend to important topics like quantum cryptography (see [18, 19]) and will be helpful in optics-based quantum information processing. We would like to thank Mr. Flammia for enlightening discussions on the topic. We acknowledge the support by the Marie Curie Excellence Grant of the EU and the Alexander von Humboldt Foundation, the National Natural Science Foundation of China and the Chinese Academy of Sciences.

This version is a rough draft. We put it onto the quantum server to share our measurement results with the scientific community. Discussion is encouraged.

**Appendix - tables:**

| ANGLES | VALUE | SIGMA |
|---|---|---|
| E(22.5,45) | 0.578065 | σ = ± 0.011966 |
| E(22.5,0) | -0.67484 | σ = ± 0.010916 |
| E(67.5,45) | 0.600959 | σ = ± 0.011296 |
| E(67.5,0) | 0.689742 | σ = ± 0.010944 |

Table 1: Angular settings of half wave plates and measured values of $E_x$ with standard deviation.

**Appendix -figures:**

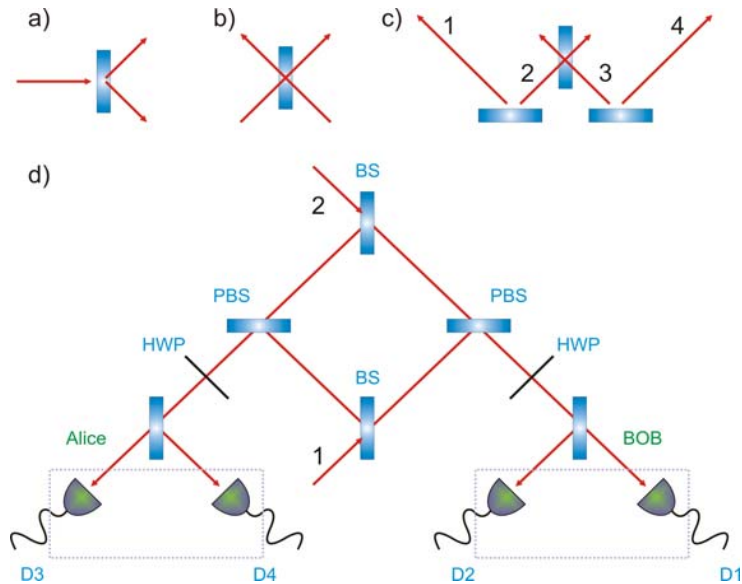

**Figure 1** Different methods for creation of entanglement. (a) A high intensity laser beam generates two entangled photons in a down conversion process. (b) Two single-photons generated independently overlap on a 50/50 beamsplitter. (c) Entanglement swapping as demonstrated by Pan et al. using two down-converted photon pairs. Photon 2 and 3 undergo a joint Bell-measurement. In consequence, Photon 1 and 4 are projected into a Bell-state despite they do not locally interact. (d) Creation of entanglement without the need for ancilla photons using our non-local method.

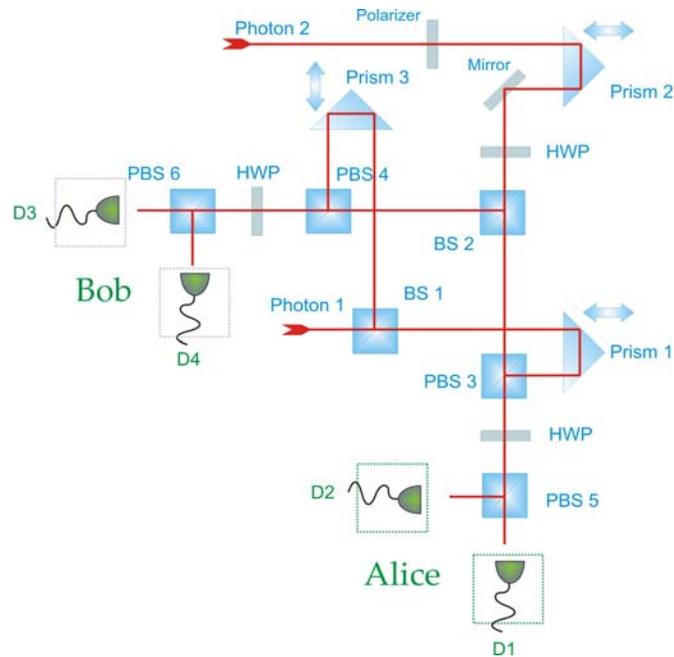

**Figure 2** Simplified schematic of the experimental setup. The two single photon enter the interferometer at the 50/50 Beamsplitters BS1 and BS2 and exit at the polarizing Beamsplitters PBS3 and PBS4. Rotating the HWP in front of PBS5 and PBS6 it is possible (a) to choose in which basis the measurement shall be performed and (b) to select polarization states that maximize the violation of the inequality of Clauser, Horne, Shimony and Holt. Successful events are registered when coincidences between opposite sites occur.

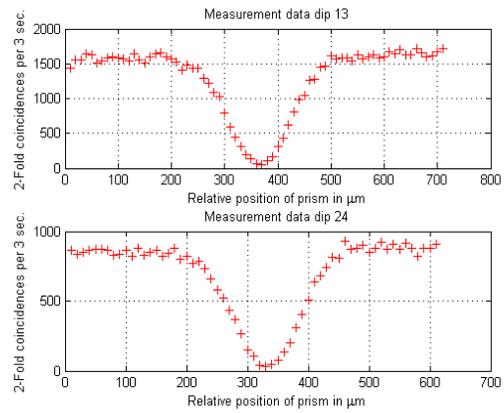

**Figure 3** a) Data of 2-4-dip (Bob) measurement to verify path length inside the interferometer. b) Data of 1-3-dip (Alice) measurement to verify path length inside the interferometer.

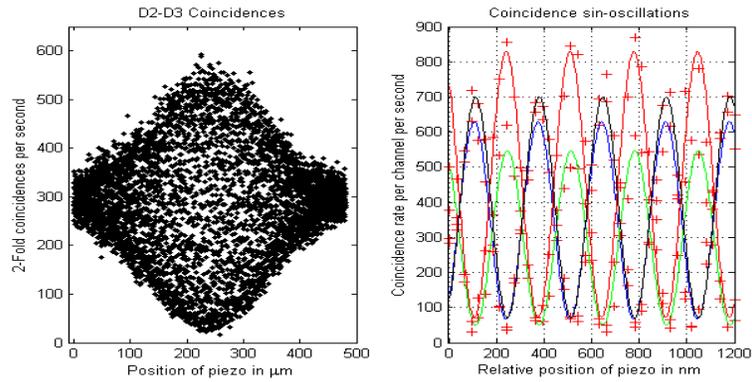

**Figure 4** (a) Measurement of the coherent envelope between detector D2 and detector D3 with visibility of 95.1. The visibilities of the other 3 combinations (D1-D2, D2-D4, D1-D4) are all above the required limit of 0.71. (b) 45-45-Measurement of relative phase change of all four coincidences dependent on piezo shift.